  \documentclass[12pt]{article}
  \usepackage{times}
  \usepackage{amsfonts}

  \topmargin - 0.5 cm 
\begin{document}

   \baselineskip = 0.80 true cm

\begin{center}
{\large \bf GENERALIZED SPIN BASES FOR QUANTUM CHEMISTRY AND QUANTUM INFORMATION\footnote{Dedicated 
to Professor Rudolf Zahradnik on the occasion of his 80th birthday.}} 
\end{center}

\vspace{0.35cm}

\begin{center}
{\bf Maurice R.~KIBLER}
\end{center}

\begin{center}
{Universit\'e de Lyon, F--69622, Lyon, France; \\
Universit\'e Lyon 1, Villeurbanne; \\
CNRS/IN2P3, UMR5822, Institut de Physique Nucl\'eaire de Lyon} \\
email: m.kibler@ipnl.in2p3.fr
\end{center}

\vspace{0.35cm}

\noindent Symmetry adapted bases in quantum chemistry and bases adapted to quantum 
information share a common characteristics: both of them are constructed 
from subspaces of the representation space of the group $SO(3)$ 
or its double group (i.e., spinor group) $SU(2)$. We exploit this fact for 
generating spin bases of relevance for quantum systems with cyclic symmetry  
and equally well for quantum information and quantum computation. 
Our approach is based on the use of generalized Pauli matrices arising 
from a polar decomposition of  $SU(2)$. This approach leads to a complete 
solution for the construction of mutually unbiased 
bases in the case where the dimension $d$ of the considered Hilbert subspace is a 
prime number. We also give the starting point for studying the case where $d$ is 
the power of a prime number. A connection of this work with the unitary group 
$U(d)$ and the Pauli group is brielly underlined.

\vspace{0.25cm}


\newpage

\section*{Introduction}

The notion of symmetry adapted functions (or vectors) in physical chemistry 
and solid state physics goes back to the fifties$^{1}$. The use of bases 
consisting of such functions allows to simplify the calculation of matrix 
elements of operators and to factorize the secular equation. Symmetry adaptation 
generally requires two type of groups: the symmetry group for the hamiltonian 
(often a finite group when dealing with molecules) and a chain of classification 
groups for the operators and state vectors (often continuous groups like
unitary groups$^{2, 3, 4}$ and finite groups$^{5, 6, 7, 8, 9}$).  The interest 
of symmetry adapted bases (atomic orbitals, molecular orbitals, spin waves, etc.) 
is well-known in quantum chemistry. In particular, the spherical harmonics 
(e.g., in atomic spectroscopy) and cubical, tetragonal or trigonal harmonics 
(e.g., in crystal-field theory and ligand field theory$^{10}$) are quite 
familiar to the practitioner in theoretical chemistry and chemical physics.

The symmetry adapted functions generally span bases for finite-dimensional Hilbert 
spaces associated with reducible or irreducible representations of a symmetry group. 
In the case of low dimensions, such spaces are especially useful in the emerging fields 
of quantum information and quantum computation (quantum state tomography and quantum 
cryptography), two fields at the crossing of informatics, mathematics and quantum 
physics. In fact, a Hilbert space of finite dimension $d$ can describe a system of 
qudits (qubits correspond to $d=2$, qudits to $d$ arbitrary). Qudits can be realized 
from many physical systems. We undersee that qudits could be also produced from 
chemical systems.

It is the object of this paper to construct bases which play an important role for 
quantum systems with cyclic symmetry and for quantum measurements and quantum information 
theory.

The organisation of this paper is as follows.  Section 1 is devoted to 
an alternative to the $\{ j^2 , j_z \}$ quantization scheme of angular 
momentum. In Section 2, this scheme is worked out for generating bases 
in a form adapted to physical and chemical cyclic systems as well as to 
quantum information. Section 3 deals with somme examples in low dimensions. 
Finally, we develop in Section 4 a systematic construction of generalized 
Pauli matrices which are at the origin of generalized spin bases. In the closing 
remarks, we mention the interest of this work for the special unitary group and 
the Pauli group.  

Throughout the present work, we use the Dirac notation familiar in quantum 
chemistry. As usual, $A^{\dagger}$ stands for the Hermitean conjugate of the 
operator $A$. In addition, $[A , B]_-$ and $[A , B]_+$ denote the commutator 
and the anticommutator of $A$ and $B$. Finally, $i$ is the pure imaginary. 

\section{AN ALTERNATIVE TO THE $\{ j^2, j_z \}$ SCHEME}
Let us consider a generalized angular momentum. We note $j^2$ its square and $j_z$ 
its $z$-component. The common eigenvectors of $j^2$ and $j_z$ are denoted as 
$| j , m \rangle$. We know that$^{11}$ 
\begin{eqnarray}
j^2 |j , m \rangle = j(j+1) |j , m \rangle, \quad j_z |j , m \rangle = m |j , m \rangle 
\end{eqnarray}
in a system of units where the rationalized Planck constant is equal to 1. For a fixed 
value of the quantum number $j$ (with $2j \in \mathbb{N}$), we note ${\cal E}(2j+1)$ 
the $(2j+1)$-dimensional Hilbert space spanned by the basis 
\begin{eqnarray}
b_s = \{ |j , m \rangle : m = j, j-1, \cdots, -j \}.
\label{spherical basis}
\end{eqnarray} 
The basis $b_s$ is adapted to spherical symmetry (adapted to the group $SO(3)$ if 
$j$ is an integer or the group $SU(2)$ if $j$ is an half on an odd integer). We take 
the basis $b_s$ in an orthonormal form, i.e., the scalar product $\langle j , m | j , m' \rangle$ 
satisfies
\begin{eqnarray}
\langle j , m | j , m' \rangle = \delta_{m , m'}
\label{scalar product jmjm'}
\end{eqnarray} 
for any value of $m$ and $m'$.

In the applications to quantum chemistry, the generalized angular momentum can be an angular 
momentum, a spin angular momentum, a total (spin $+$ orbital) angular momentum, etc. The  
vectors $|j , m \rangle$ can have several realizations. For instance, in the spectroscopy of 
$4f^N$ lanthanide ions, we have state vectors of type $|J , M \rangle \equiv |4f^N \tau S L J M \rangle$  
in the Russell-Saunders coupling (here $j = J$ and $m = M$). This constitutes one of many possible 
realizations of the vectors $|j , m \rangle$. 

Besides the basis $b_s$, another interesting basis can be obtained as follows. Let us consider the operator 
          \begin{eqnarray}
          v_{ra} = {e}^{{i} 2 \pi j r} |j , -j \rangle \langle j , j| 
                  + \sum_{m = -j}^{j-1} q^{(j-m)a} |j , m+1 \rangle \langle j , m| 
          \label{definition of vra} 
          \end{eqnarray}
where we use the notation of Dirac for projectors. In Eq. (\ref{definition of vra}), we have 
          \begin{eqnarray}
          r \in \mathbb{R}, \quad a = 0, 1, \cdots, 2j, \quad q = \exp \left( {2 \pi {i} \over 2j+1} \right). 
          \label{parameters} 
          \end{eqnarray}
The operator $v_{ra}$ is an extension of the operator$U_r$ defined in a previous work$^{12}$ 
($U_r = v_{r0}$). From Eq. (\ref{definition of vra}), we can check that the action of $v_{ra}$ on the state 
$| j , m \rangle$ is given by 
          \begin{eqnarray}
  v_{ra} |j , m \rangle = \left( 1 - \delta_{m,j} \right) q^{(j-m)a}
  |j , m+1 \rangle + \delta_{m,j}  
  {e}^{{i} 2 \pi j r}
  |j , -j \rangle. 
          \label{action of vra on jm} 
          \end{eqnarray}
Furthermore, the matrix $V_{ra}$ of the operator $v_{ra}$ on the basis $b_s$ reads   
        \begin{eqnarray}
V_{ra} = 
\pmatrix{
0                    &    q^a &      0  & \cdots &       0 \cr
0                    &      0 & q^{2a}  & \cdots &       0 \cr
\vdots               & \vdots & \vdots  & \cdots &  \vdots \cr
0                    &      0 &      0  & \cdots & q^{2ja} \cr
{e}^{{i} 2 \pi j r}  &      0 &      0  & \cdots &   0     \cr
}
        \label{definition of Vra}
        \end{eqnarray}
where the lines and columns are labeled in the order 
$|j , j \rangle, |j , j-1 \rangle, \cdots, |j , -j \rangle$. 
It can be shown that the operator $j^2$ and $v_{ra}$ commute 
so that the complete set $\{ j^2, v_{ra} \}$ of commuting 
operators constitutes an alternative to the set $\{ j^2, j_z \}$. 

We may ask what are the analogues of the vectors $|j , m \rangle$ 
in the scheme $\{ j^2, v_{ra} \}$? Indeed, they 
are the common eigenvectors of the operators $j^2$ and $v_{ra}$. As 
a result, these eigenvectors are 
          \begin{eqnarray}
|j \alpha ; r a \rangle = \frac{1}{\sqrt{2j+1}} \sum_{m = -j}^{j} 
q^{(j + m)(j - m + 1)a / 2 - j m r + (j + m)\alpha} | j , m \rangle
          \label{j alpha r a in terms of jm}, \quad \alpha = 0, 1, \cdots, 2j
          \end{eqnarray} 
for $\alpha = 0, 1, \cdots, 2j$. More precisely, we have the eigenvalue equations 
\begin{eqnarray}
v_{ra} |j \alpha ; r a \rangle = q^{j(a+r) - \alpha} |j \alpha ; r a \rangle, \quad 
j^2 |j \alpha ; r a \rangle = j(j+1) |j \alpha ; r a \rangle. 
\end{eqnarray}
For fixed $j$ and $a$ ($2j \in \mathbb{N}, a = 0, 1, \cdots, 2j$), the basis
\begin{eqnarray}
B_{ra} = \{ |j \alpha ; r a \rangle : \alpha = 0, 1, \cdots, 2j \}
\label{basis Bra}
\end{eqnarray} 
is an orthonormal basis since
\begin{eqnarray}
\langle j \alpha ; r a | j \alpha' ; r a \rangle = \delta_{\alpha , \alpha'}
\label{scalar product alphaalpha'}
\end{eqnarray} 
for any value of $\alpha$ and $\alpha'$. In the particular case where $2j+1$ is a prime integer, 
the overlap between the bases $B_{ra}$ and $B_{rb}$ is such that$^{13}$
          \begin{eqnarray}
| \langle j \alpha ; r a | j \beta ; r b \rangle | = 
\delta_{\alpha , \beta} \delta_{a , b} + \frac{1}{\sqrt{2j+1}} (1 - \delta_{a , b}) 
          \label{mub relation}
          \end{eqnarray} 
a property of considerable importance in quantum information. Note that Eq. (\ref{mub relation}) 
is compatible with Eq. (\ref{scalar product alphaalpha'}).

\section{A FORMULATION FOR $d$-DIMENSIONAL QUANTUM SYSTEMS}
The parameter $r$ is of interest for group-theoretical analyses but turns out to be of no concern 
here. Therefore, we shall restrict ourselves in the following to the case $r=0$. In addition, we shall 
adopt the notation 
          \begin{eqnarray}
k = j - m, \quad | k \rangle = | j , m \rangle, \quad | a \alpha \rangle = | j \alpha ; 0 a \rangle, \quad d = 2j+1
          \label{notation QIT}
          \end{eqnarray}  
that is especially adapted to quantum information (the vectors 
$| 0 \rangle, | 1 \rangle, \cdots, | d-1 \rangle$ are then called 
qudits, the case $d=2$ corresponding to ordinary qubits) and to cyclic chemical systems (for which 
$| d \rangle \equiv | 0 \rangle$, $| d+1 \rangle \equiv | 1 \rangle$, \ldots). 

The basis $b_s$ becomes 
	\begin{eqnarray}
B_d = \{ | k \rangle : k = 0, 1, \cdots, d-1 \}
	\label{computational basis}
	\end{eqnarray} 
known as the computational basis in quantum information theory. The action of $v_{ra}$ on the basis $B_d$ 
of ${\cal E}(2j+1)$ is described by 
          \begin{eqnarray}
  v_{0a} | k \rangle = q^{ka} | k-1 \rangle 
          \label{action of v0a on k} 
          \end{eqnarray}
where $k-1$ should be understood modulo $d$ (i.e., $| -1 \rangle = | d -1 \rangle$). The vectors $| a \alpha \rangle$ 
of the orthonormal basis 
\begin{eqnarray}
B_{ra} = \{ | a \alpha \rangle : \alpha = 0, 1, \cdots, d-1 \}
\label{basis B0a}
\end{eqnarray} 
can be written as 
          \begin{eqnarray}
| a \alpha \rangle = \frac{1}{\sqrt{d}} \sum_{k = 0}^{d-1} 
q^{(d - k - 1)(k + 1)a / 2 - (k + 1) \alpha} | k \rangle
          \label{a alpha in terms of k}
          \end{eqnarray} 
where $\alpha$ can take the values $\alpha = 0, 1, \cdots, d-1$. These vectors satisfy the eigenvalue equation
	  \begin{eqnarray}
v_{0a} | a \alpha \rangle = q^{(d-1)a / 2 - \alpha} | a \alpha \rangle 
        \label{eigenvalue equation}
        \end{eqnarray}
that corresponds non a nondegenerate spectrum for the operator $v_{0a}$. 

All relations given in Section 1 up to this point are valid for $d$ arbitrary. In the 
special case where $d$ is a prime integer, Eq. (\ref{mub relation}) yields 
	    \begin{eqnarray}
| \langle a \alpha | b \beta \rangle | = 
\delta_{\alpha , \beta} \delta_{a , b} + \frac{1}{\sqrt{d}} (1 - \delta_{a , b}) 
          \label{mub relation in terms of a-alpha}
          \end{eqnarray} 
a relation valid for any value of $a$, $b$, $\alpha$ and $\beta$ in the set $\{ 0, 1, \cdots, d-1 \}$.
In quantum information, two bases $B_{0a}$ and $B_{0b}$ satisfying Eq. (\ref{mub relation in terms of a-alpha}) 
are said to be mutually unbiased$^{14}$. Such bases play an important role in quantum 
cryptography and quantum state tomography. It is well-known that a complete set of $d+1$ mutually 
unbiased bases can be found when $d$ is a prime integer or the power of a prime integer. 

We continue with some typical examples.

\section{SOME TYPICAL EXAMPLES}
\subsection*{The case $d=2$}
In this case, relevant for a spin $j = 1/2$ or for a qubit, we have $q = -1$ and $a, \alpha \in \{ 0 , 1 \}$. The matrices of the 
operators $v_{0a}$ are 
     \begin{eqnarray}
V_{00} = 
\pmatrix{
  0     &1   \cr
  1     &0   \cr
}, \quad 
V_{01} = 
\pmatrix{
  0     &-1  \cr
  1     &0   \cr
}.
     \end{eqnarray}
We note in passing a connection (to be generalized below) with the Pauli matrices since 
$V_{00} = \sigma_x$ and $V_{01} = -i \sigma_y$. From Eqs. (\ref{computational basis}), 
(\ref{basis B0a}) and (\ref{a alpha in terms of k}), 
the bases $B_{2}$, $B_{00}$ and $B_{01}$ are 
     \begin{eqnarray}
B_{2}  & : & | 0 \rangle, \quad 
           | 1 \rangle 
	\label{1cas d is 2} \\
B_{00} &:& | 0 0 \rangle = \frac{1}{\sqrt{2}} \left(    | 0 \rangle + | 1 \rangle \right), \quad
           | 0 1 \rangle = \frac{1}{\sqrt{2}} \left( -  | 0 \rangle + | 1 \rangle \right)  
	\label{2cas d is 2} \\
B_{01} &:& | 1 0 \rangle = \frac{1}{\sqrt{2}} \left(  i | 0 \rangle + | 1 \rangle \right), \quad
           | 1 1 \rangle = \frac{1}{\sqrt{2}} \left( -i | 0 \rangle + | 1 \rangle \right) 
	\label{3cas d is 2}
     \end{eqnarray}
which satisfy Eq. (\ref{mub relation in terms of a-alpha}). Note that by using the spinorbital 
	\begin{eqnarray}
\alpha = | \frac{1}{2} ,   \frac{1}{2} \rangle = | 0 \rangle, \quad 
\beta  = | \frac{1}{2} , - \frac{1}{2} \rangle = | 1 \rangle 
     \end{eqnarray}
($\alpha$ for spin up and $\beta$ for spin down) familiar to the quantum chemist, 
Eqs. (\ref{1cas d is 2})-(\ref{3cas d is 2}) can be rewritten as 
	\begin{eqnarray}
B_{2}  &:& \alpha, \quad
           \beta \\
B_{00} &:& | 0 0 \rangle =    \frac{1}{\sqrt{2}} \left( \alpha +   \beta \right), \quad
           | 0 1 \rangle = -  \frac{1}{\sqrt{2}} \left( \alpha -   \beta \right)  \\
B_{01} &:& | 1 0 \rangle =  i \frac{1}{\sqrt{2}} \left( \alpha - i \beta \right), \quad
           | 1 1 \rangle = -i \frac{1}{\sqrt{2}} \left( \alpha + i \beta \right). 
     \label{cas d is 2 in alpha and beta}
     \end{eqnarray}
In terms of eigenvectors of the matrices $V_{0a}$, we must replace the vectors $| a \alpha \rangle$ 
by column vectors. This leads to
	\begin{eqnarray}
B_{2}  &:& \alpha \to 
\pmatrix{
  1  \cr
  0  \cr
}, \quad
           \beta  \to 
\pmatrix{
  0  \cr
  1  \cr
} \\
B_{00} &:& | 0 0 \rangle \to    \frac{1}{\sqrt{2}} \pmatrix{
  1  \cr
  1  \cr
}, \quad
           | 0 1 \rangle \to -  \frac{1}{\sqrt{2}} \pmatrix{
  1  \cr
 -1  \cr
} \\
B_{01} &:& | 1 0 \rangle \to  i \frac{1}{\sqrt{2}} \pmatrix{
  1  \cr
 -i  \cr
}, \quad
           | 1 1 \rangle \to -i \frac{1}{\sqrt{2}} \pmatrix{
  1  \cr
  i  \cr
}. 
     \label{cas d is 2 column vectors}
     \end{eqnarray}

\subsection*{The case $d=3$}
This case corresponds to a spin $j=1$ or to a qutrit. Here, we have 
$q = \exp (2 \pi i / 3)$ and $a, \alpha \in \{ 0 , 1 , 2 \}$. The 
matrices of the operators $v_{0a}$ are 
	\begin{eqnarray}
V_{00} = 
\pmatrix{
  0     &1   &0 \cr
  0     &0   &1 \cr
  1     &0   &0 \cr
}, \quad 
\pmatrix{
  0     &q   &0   \cr
  0     &0   &q^2 \cr
  1     &0   &0   \cr
}, \quad
\pmatrix{
  0     &q^2   &0 \cr
  0     &0     &q \cr
  1     &0     &0 \cr
}.
	\end{eqnarray}
The bases $B_{3}$, $B_{00}$ and $B_{01}$ $B_{02}$ are
     \begin{eqnarray}
B_{3}:  & & | 0 \rangle,  \ 
            | 1 \rangle,  \  
            | 2 \rangle   \\
B_{00}: & & | 0 0 \rangle = \frac{1}{\sqrt{3}} \left(     | 0 \rangle +     | 1 \rangle + | 2 \rangle \right),  \ 
            | 0 1 \rangle = \frac{1}{\sqrt{3}} \left( q^2 | 0 \rangle + q   | 1 \rangle + | 2 \rangle \right)   \\ 
        & & | 0 2 \rangle = \frac{1}{\sqrt{3}} \left( q   | 0 \rangle + q^2 | 1 \rangle + | 2 \rangle \right)   \\
B_{01}: & & | 1 0 \rangle = \frac{1}{\sqrt{3}} \left( q   | 0 \rangle + q   | 1 \rangle + | 2 \rangle \right),  \ 
            | 1 1 \rangle = \frac{1}{\sqrt{3}} \left(     | 0 \rangle + q^2 | 1 \rangle + | 2 \rangle \right)   \\ 
        & & | 1 2 \rangle = \frac{1}{\sqrt{3}} \left( q^2 | 0 \rangle +     | 1 \rangle + | 2 \rangle \right)   \\
B_{02}: & & | 2 0 \rangle = \frac{1}{\sqrt{3}} \left( q^2 | 0 \rangle + q^2 | 1 \rangle + | 2 \rangle \right), \
            | 2 1 \rangle = \frac{1}{\sqrt{3}} \left( q   | 0 \rangle +     | 1 \rangle + | 2 \rangle \right)   \\ 
        & & | 2 2 \rangle = \frac{1}{\sqrt{3}} \left(     | 0 \rangle + q   | 1 \rangle + | 2 \rangle \right). 
     \label{cas d is 3}
     \end{eqnarray}
They satisfy Eq. (\ref{mub relation in terms of a-alpha}). In terms of colum vectors, we have 
\begin{eqnarray}
B_{3}  &:& | 0 \rangle \to 
\pmatrix{
  1  \cr
  0  \cr
  0  \cr
}, \quad
           | 1 \rangle \to 
\pmatrix{
  0  \cr
  1  \cr
  0  \cr
}, \quad
           | 2 \rangle \to 
\pmatrix{
  0  \cr
  0  \cr
  1  \cr
} \\
B_{00} &:& | 0 0 \rangle \to    \frac{1}{\sqrt{3}} \pmatrix{
  1  \cr
  1  \cr
  1  \cr
}, \quad
           | 0 1 \rangle \to   \frac{1}{\sqrt{3}} \pmatrix{
  q^2  \cr
  q    \cr
  1    \cr
}, \quad
           | 0 2 \rangle \to   \frac{1}{\sqrt{3}} \pmatrix{
  q   \cr
  q^2 \cr
  1   \cr
} \\
B_{01} &:& | 1 0 \rangle \to    \frac{1}{\sqrt{3}} \pmatrix{
  q  \cr
  q  \cr
  1  \cr
}, \quad
           | 1 1 \rangle \to   \frac{1}{\sqrt{3}} \pmatrix{
  1    \cr
  q^2  \cr
  1    \cr
}, \quad
           | 1 2 \rangle \to   \frac{1}{\sqrt{3}} \pmatrix{
  q^2    \cr
  1      \cr
  1      \cr
} \\
B_{02} &:& | 2 0 \rangle \to    \frac{1}{\sqrt{3}} \pmatrix{
  q^2  \cr
  q^2  \cr
  1  \cr
}, \quad
           | 2 1 \rangle \to   \frac{1}{\sqrt{3}} \pmatrix{
  q  \cr
  1  \cr
  1  \cr
}, \quad
           | 2 2 \rangle \to   \frac{1}{\sqrt{3}} \pmatrix{
  1  \cr
  q  \cr
  1  \cr
}. 
     \label{cas d is 3 column vectors}
     \end{eqnarray}

\subsection*{The case $d=4$}
This case corresponds to a spin $j = 3/2$. Here, we have $q = i$ and $a, \alpha \in \{ 0 , 1 , 2, 3 \}$. Equation 
(\ref{a alpha in terms of k}) can be applied to this case too. However, the resulting bases 
$B_4$, $B_{00}$, $B_{01}$, $B_{02}$ and $B_{03}$ do not constitute a complete system of mutually unbiased 
bases ($d=4$ is not a prime number). Nevertheless, it is possible to find $d+1 = 5$ mutually unbiased bases 
because $d = 2^2$ is the power of a prime number. This can be achieved by replacing the space ${\cal E}(4)$ 
spanned by $\{ | 3/2 , m \rangle : m = 3/2, 1/2, -1/2, -3/2 \}$ by the tensor product space
${\cal E}(2) \otimes {\cal E}(2)$ spanned by the basis 
	\begin{eqnarray}
\{ \alpha \otimes \alpha, \alpha \otimes \beta, \beta \otimes \alpha, \beta \otimes \beta \}.
	\label{base pd}
	\end{eqnarray}
The space ${\cal E}(2) \otimes {\cal E}(2)$ is associated with the coupling of two spin angular momenta 
$j_1 = 1/2$ and $j_2 = 1/2$ or two qubits (in the vector $u \otimes v$, $u$ and $v$ correspond to 
$j_1$       and $j_2$, respectively). An alternative basis for ${\cal E}(2) \otimes {\cal E}(2)$ is 
	\begin{eqnarray}
\{ \alpha \otimes \alpha, \frac{1}{2} (\alpha \otimes \beta + \beta \otimes \alpha), 
   \beta  \otimes \beta,  \frac{1}{2} (\alpha \otimes \beta - \beta \otimes \alpha) \}.
	\label{base pd SU2 S2}
	\end{eqnarray}
The vectors in (\ref{base pd}) are well-known in the treatment of spin systems. The first three vectors are 
symmetric under the interchange $1 \leftrightarrow 2$ and describe a total angular momentum $J=1$ while the last 
one is antisymmetric and corresponds to $J=0$. It should be observed that the basis (\ref{base pd SU2 S2}) 
illustrates a connection between the special unitary group $SU(2)$ and the permutation group $S_2$ (a particular 
case of a reciprocity theorem between irreducible representation classes of $SU_n$ and $S_m$).  

In addition to the bases (\ref{base pd}) and (\ref{base pd SU2 S2}), it is possible to find other bases of 
${\cal E}(2) \otimes {\cal E}(2)$ which are mutually unbiased. The $d=4$ mutually unbiased bases besides the 
canonical or computational basis (\ref{base pd}) can be constructed from the eigenvectors 
	\begin{eqnarray}
|a b \alpha \beta \rangle = |a \alpha \rangle \otimes |b \beta \rangle 
	\label{tensor product of vectors}
	\end{eqnarray}
of the operators $w_{ab} = v_{0a} \otimes v_{0b}$ (the vectors $|a \alpha \rangle$ and $|b \beta \rangle$ refer 
to the two spaces ${\cal E}(2)$). As a result, we have the $d+1 = 5$ following mutually unbiased bases where 
$\lambda = (1-i)/2$ and $\mu = (1+i)/2$.

\noindent {\bf The canonical basis:}
	\begin{eqnarray}
\alpha \otimes \alpha, \quad \alpha \otimes \beta, \quad \beta \otimes \alpha, \quad \beta \otimes \beta
	\label{canonical basis}
	\end{eqnarray}
or in column vectors
	\begin{eqnarray}
\pmatrix{
1 \cr
0 \cr
0 \cr
0 \cr
}, \quad
\pmatrix{
0 \cr
1 \cr
0 \cr
0 \cr
}, \quad
\pmatrix{
0 \cr
0 \cr
1 \cr
0 \cr
}, \quad
\pmatrix{
0 \cr
0 \cr
0 \cr
1 \cr
}.
	\label{canonical basis}
	\end{eqnarray}
\noindent {\bf The $w_{00}$ basis:}
	\begin{eqnarray}
| 0 0 0 0 \rangle &=& \frac{1}{2} 
(\alpha \otimes \alpha + \alpha \otimes \beta + \beta \otimes \alpha + \beta \otimes \beta) \\
| 0 0 0 1 \rangle &=& \frac{1}{2} 
(\alpha \otimes \alpha - \alpha \otimes \beta + \beta \otimes \alpha - \beta \otimes \beta) \\
| 0 0 1 0 \rangle &=& \frac{1}{2} 
(\alpha \otimes \alpha + \alpha \otimes \beta - \beta \otimes \alpha - \beta \otimes \beta) \\
| 0 0 1 1 \rangle &=& \frac{1}{2} 
(\alpha \otimes \alpha - \alpha \otimes \beta - \beta \otimes \alpha + \beta \otimes \beta)
	\label{w00 basis}
	\end{eqnarray}
or in column vectors
	\begin{eqnarray}
\frac{1}{2} \pmatrix{
1 \cr
1 \cr
1 \cr
1 \cr
}, \quad
\frac{1}{2} \pmatrix{
1 \cr
-1 \cr
1 \cr
-1 \cr
}, \quad
\frac{1}{2} \pmatrix{
1 \cr
1 \cr
-1 \cr
-1 \cr
}, \quad
\frac{1}{2} \pmatrix{
1 \cr
-1 \cr
-1 \cr
1 \cr
}.
	\label{canonical basis}
	\end{eqnarray}
\noindent {\bf The $w_{11}$ basis:}
	\begin{eqnarray}
| 1 1 0 0 \rangle &=& \frac{1}{2} 
(\alpha \otimes \alpha + i \alpha \otimes \beta + i \beta \otimes \alpha - \beta \otimes \beta) \\
| 1 1 0 1 \rangle &=& \frac{1}{2} 
(\alpha \otimes \alpha - i \alpha \otimes \beta + i \beta \otimes \alpha + \beta \otimes \beta) \\
| 1 1 1 0 \rangle &=& \frac{1}{2} 
(\alpha \otimes \alpha + i \alpha \otimes \beta - i \beta \otimes \alpha + \beta \otimes \beta) \\
| 1 1 1 1 \rangle &=& \frac{1}{2} 
(\alpha \otimes \alpha - i \alpha \otimes \beta - i \beta \otimes \alpha - \beta \otimes \beta)
	\label{w11 basis}
	\end{eqnarray}
or in column vectors
	\begin{eqnarray}
\frac{1}{2} \pmatrix{
1 \cr
i \cr
i \cr
-1 \cr
}, \quad
\frac{1}{2} \pmatrix{
1 \cr
-i \cr
i \cr
1 \cr
}, \quad
\frac{1}{2} \pmatrix{
1 \cr
i \cr
-i \cr
1 \cr
}, \quad
\frac{1}{2} \pmatrix{
1 \cr
-i \cr
-i \cr
-1 \cr
}.
	\label{canonical basis}
	\end{eqnarray}
\noindent {\bf The $w_{01}$ basis:}
	\begin{eqnarray}
\lambda | 0 1 0 0 \rangle + \mu | 0 1 1 1 \rangle &=& \frac{1}{2} 
(\alpha \otimes \alpha + \alpha \otimes \beta - i \beta \otimes \alpha + i \beta \otimes \beta) \\
\mu | 0 1 0 0 \rangle + \lambda | 0 1 1 1 \rangle &=& \frac{1}{2} 
(\alpha \otimes \alpha - \alpha \otimes \beta + i \beta \otimes \alpha + i \beta \otimes \beta) \\
\lambda | 0 1 0 1 \rangle + \mu | 0 1 1 0 \rangle &=& \frac{1}{2} 
(\alpha \otimes \alpha - \alpha \otimes \beta - i \beta \otimes \alpha - i \beta \otimes \beta) \\
\mu | 0 1 0 1 \rangle + \lambda | 0 1 1 0 \rangle &=& \frac{1}{2} 
(\alpha \otimes \alpha + \alpha \otimes \beta + i \beta \otimes \alpha - i \beta \otimes \beta)
	\label{w01 basis}
	\end{eqnarray}
or in column vectors
	\begin{eqnarray}
\frac{1}{2} \pmatrix{
1 \cr
1 \cr
-i \cr
i \cr
}, \quad
\frac{1}{2} \pmatrix{
1 \cr
-1 \cr
i \cr
i \cr
}, \quad
\frac{1}{2} \pmatrix{
1 \cr
-1 \cr
-i \cr
-i \cr
}, \quad
\frac{1}{2} \pmatrix{
1 \cr
1 \cr
i \cr
-i \cr
}.
	\label{canonical basis}
	\end{eqnarray}
\noindent {\bf The $w_{10}$ basis:}
	\begin{eqnarray}
\lambda | 1 0 0 0 \rangle + \mu | 1 0 1 1 \rangle &=& \frac{1}{2} 
(\alpha \otimes \alpha - i \alpha \otimes \beta + \beta \otimes \alpha + i \beta \otimes \beta) \\
\mu | 1 0 0 0 \rangle + \lambda | 1 0 1 1 \rangle &=& \frac{1}{2} 
(\alpha \otimes \alpha + i \alpha \otimes \beta - \beta \otimes \alpha + i \beta \otimes \beta) \\
\lambda | 1 0 0 1 \rangle + \mu | 1 0 1 0 \rangle &=& \frac{1}{2} 
(\alpha \otimes \alpha + i \alpha \otimes \beta + \beta \otimes \alpha - i \beta \otimes \beta) \\
\mu | 1 0 0 1 \rangle + \lambda | 1 0 1 0 \rangle &=& \frac{1}{2} 
(\alpha \otimes \alpha - i \alpha \otimes \beta - \beta \otimes \alpha - i \beta \otimes \beta)
	\label{w10 basis}
	\end{eqnarray}
or in column vectors
	\begin{eqnarray}
\frac{1}{2} \pmatrix{
1 \cr
-i \cr
1 \cr
i \cr
}, \quad
\frac{1}{2} \pmatrix{
1 \cr
i \cr
-1 \cr
i \cr
}, \quad
\frac{1}{2} \pmatrix{
1 \cr
i \cr
1 \cr
-i \cr
}, \quad
\frac{1}{2} \pmatrix{
1 \cr
-i \cr
-1 \cr
-i \cr
}.
	\label{canonical basis}
	\end{eqnarray}
It is to be noted that the vectors of the $w_{00}$ and $w_{11}$ bases are not intricated 
(i.e., each vector is the direct product of two vectors) while the vectors of the $w_{01}$ 
and $w_{10}$ bases are intricated (i.e., each vector is not the direct product of two vectors). 

\section{GENERALIZED PAULI MATRICES}
From the operators $v_{0a}$, it is possible to define two basic operators $x$ and $z$ which can be 
used for generating generalized Pauli matrices. Let us put 
	\begin{eqnarray}
x = v_{00}, \quad z = v_{00}^{\dagger} v_{01}. 
	\label{definition of x and z}
	\end{eqnarray}
The action of $x$ and $z$ on the space ${\cal E}(2j+1)$ is given by 
	\begin{eqnarray}
  x |j , m \rangle = \left( 1 - \delta_{m,j} \right) |j , m+1 \rangle + \delta_{m,j} |j , -j \rangle 
\Leftrightarrow  x | k \rangle = | k-1 \rangle 
	\label{action of x on jm} 
	\end{eqnarray}
and 
          \begin{eqnarray}
z | j,m \rangle = q^{j-m} | j,m \rangle \Leftrightarrow  z | k \rangle = q^{k} | k \rangle
          \label{action de z sur jm} 
          \end{eqnarray}
where $q = \exp (2 \pi i / d)$ with $d = 2j+1$. The $d$-dimensional matrices $X$ and $Z$ of $x$ and $z$ are
        \begin{eqnarray}
X = 
\pmatrix{
0                    &      1 &      0  & \cdots &       0 \cr
0                    &      0 &      1  & \cdots &       0 \cr
\vdots               & \vdots & \vdots  & \cdots &  \vdots \cr
0                    &      0 &      0  & \cdots &       1 \cr
1                    &      0 &      0  & \cdots &       0 \cr
}, \quad
Z = 
\pmatrix{
1                    &      0 &      0    & \cdots &       0       \cr
0                    &      q &      0    & \cdots &       0       \cr
0                    &      0 &      q^2  & \cdots &       0       \cr
\vdots               & \vdots & \vdots    & \cdots &  \vdots       \cr
1                    &      0 &      0    & \cdots &       q^{d-1} \cr
}.
        \label{definition of X and Z}
        \end{eqnarray}
The operators $x$ and $z$ are unitary and satisfy the $q$-commutation relation
          \begin{eqnarray}
x z - q z x = 0.
          \label{q commutation} 
          \end{eqnarray}
Pairs of operators satisfying a relation of type (\ref{q commutation}) are 
nowadays refered to as Weyl pairs. Thus, the operators $x$ and $z$ constitute 
a Weyl pair. Weyl pairs were introduced at the begining of quantum mechanics$^{15}$. 
They were extensively used for factorizing the secular equation in connection with 
a study of alternating hydrocarbons$^{16}$ and for constructing analogues of the 
usual Pauli matrices$^{14, 17, 18, 19}$.

Let us now define the operators 
     \begin{eqnarray}
u_{ab} = x^a z^b, \quad a, b = 0, 1, \cdots, d-1. 
     \end{eqnarray}
The $d^2$ operators $u_{ab}$ are unitary and satisfy the following trace relation 
          \begin{eqnarray}
 {\rm Tr}_{{\cal E}(2j+1)} \left( u_{ab}^{\dagger} u_{a'b'} \right) = 
 (2j+1) \>
 \delta_{a,a'} \> 
 \delta_{b,b'} 
          \label{trace de uu}
          \end{eqnarray}
where the trace is taken on the $d$-dimensional space ${\cal E}(2j+1)$. Additionally, the commutator 
                $[u_{ab} , u_{a'b'}]_-$ and the 
anti-commutator $[u_{ab} , u_{a'b'}]_+$ of $u_{ab}$ and $u_{a'b'}$ are given by 
          \begin{eqnarray}
[u_{ab} , u_{a'b'}]_{\mp} = \left( q^{-ba'} \mp q^{-ab'} \right) u_{a'' b''}, 
\quad a'' = a+a', 
\quad b'' = b+b'. 
          \label{com anti-com}
          \end{eqnarray}
Consequently, $[u_{ab} , u_{a'b'}]_{-} = 0$ if and only if $ab' - ba' = 0$
(mod $d$) and $[u_{ab} , u_{a'b'}]_{+} = 0$ if and only if $ab' - ba' = (1/2) d$
(mod $d$). Therefore, all anti-commutators $[u_{ab} , u_{a'b'}]_{+}$ are
different from 0 if $d$ is an odd integer.  

Two consequences follow from Eqs. (\ref{trace de uu}) and (\ref{com anti-com}). First, the trace relation 
(\ref{trace de uu}) shows that the $d^2$ operators $u_{ab}$ are pairwise orthogonal operators so that they 
can serve as a basis for developing any operator acting on the Hilbert space ${\cal E}(d)$. Second, the 
commutation relation (\ref{com anti-com}) shows that the set $\{ u_{ab} : a, b = 0, 1, \cdots, d-1 \}$ 
generates a $d^2$-dimensional Lie algebra. This algebra turns out to be the Lie algebra of the unitary group 
$U(d)$. The subset $\{ u_{ab} : a, b = 0, 1, \cdots, d-1 \} \setminus \{ u_{00} \}$ thus spans the Lie algebra 
of the special unitary group $SU(d)$. 

All this is reminiscent of the group $SU(2)$, the generators of which are the well-known Pauli matrices. 
Therefore, the operators $u_{ab}$ shall be refered as generalized Pauli operators and their matrices as 
generalized Pauli matrices. As an illustration, let us deal with the cases $d=2$ and $d=3$.

\subsection*{Exemple 1}
In the case $j = 1/2 \Leftrightarrow d = 2$ ($\Rightarrow q = -1$), the matrices of the 4 
operators $u_{ab}$ with $a, b = 0,1$ are 
     \begin{eqnarray}
I = X^0 Z^0 = 
\pmatrix{
  1     &0   \cr
  0     &1   \cr
}, \quad 
X = X^1 Z^0 = 
\pmatrix{
  0     &1   \cr
  1     &0   \cr
}
     \end{eqnarray}
     \begin{eqnarray}
Z = X^0 Z^1 = 
\pmatrix{
  1     &0   \cr
  0     &-1  \cr
}, \quad 
Y = X^1 Z^1 = 
\pmatrix{
  0     &-1  \cr
  1     &0   \cr
}.
     \end{eqnarray}
In terms of the usual (Hermitean and unitary) Pauli matrices 
$\sigma_x$, $\sigma_y$ and $\sigma_z$, we have $X = \sigma_x$,   
$Y = - i \sigma_y$, $Z = \sigma_z$. The approach 
developed in the present paper lead to Pauli matrices in dimension 2 
that differ from the usual Pauli matrices. This is the price one has to pay in order 
to get a systematic generalization of Pauli matrices in arbitrary dimension. 
It should be observed that the commutation and anti-commutation 
relations given by (\ref{com anti-com}) with $d=2$ correspond to the well-known 
commutation and anti-commutation relations for the usual Pauli matrices (transcribed 
in the normalization 
$X^1 Z^0 = \sigma_x$, 
$X^1 Z^1 = -i \sigma_y$, 
$X^0 Z^1 = \sigma_z$).

\subsection*{Exemple 2}
In the case $j = 1 \Leftrightarrow d = 3$ ($\Rightarrow q = \exp(2 \pi i/3)$), the matrices of the 9  
operators $u_{ab}$ with $a, b = 0,1,2$, viz., 
     \begin{eqnarray}
X^0 Z^0 = I      \quad 
X^1 Z^0 = X      \quad
X^2 Z^0 = X^2    \quad
X^0 Z^1 = Z      \quad
X^0 Z^2 = Z^2    
     \end{eqnarray}
     \begin{eqnarray}
X^1 Z^1 = X Z    \quad 
X^2 Z^2          \quad
X^2 Z^1 = X^2 Z  \quad  
X^1 Z^2 = X Z^2  
     \end{eqnarray}
are 
     \begin{eqnarray}
I = 
\pmatrix{
  1     &0     &0   \cr
  0     &1     &0   \cr
  0     &0     &1   \cr
}, \quad 
X = 
\pmatrix{
  0     &1     &0   \cr
  0     &0     &1   \cr
  1     &0     &0   \cr
}, \quad 
X^2 = 
\pmatrix{
  0     &0     &1   \cr
  1     &0     &0   \cr
  0     &1     &0   \cr
}
     \end{eqnarray}
     \begin{eqnarray}
Z = 
\pmatrix{
  1     &0     &0     \cr
  0     &q     &0     \cr
  0     &0     &q^2   \cr
}, \quad 
Z^2 = 
\pmatrix{
  1     &0       &0   \cr
  0     &q^2     &0   \cr
  0     &0       &q   \cr
}, \quad 
X Z = 
\pmatrix{
  0     &q     &0     \cr
  0     &0     &q^2   \cr
  1     &0     &0     \cr
}
     \end{eqnarray}
     \begin{eqnarray}
X^2 Z^2 = 
\pmatrix{
  0     &0       &q     \cr
  1     &0       &0     \cr
  0     &q^2     &0     \cr
}, \quad 
X^2 Z =  
\pmatrix{
  0     &0     &q^2     \cr
  1     &0     &0       \cr
  0     &q     &0       \cr
}, \quad 
X Z^2 =  
\pmatrix{
  0     &q^2     &0     \cr
  0     &0       &q     \cr
  1     &0       &0     \cr
}.
     \end{eqnarray}
These generalized Pauli matrices differ from the Gell-Mann matrices$^{20}$ used in elementary 
particle physics. They constitute a natural extension of the Pauli matrices in dimension $d = 3$ .

\section{CONCLUDING REMARKS}
The various bases described in the present paper are of central importance in quantum information and 
quantum computation. They also play an important role for quantum (chemical and physical) systems with 
cyclic symmetry. By way of illustration, we would like to mention two examples.
 
Let us consider a ring shape molecule with $N$ atoms (or agregates) 
at the vertices of a regular polygon with $N$ sides ($N=6$ for 
the benzen molecule C$_6$H$_6$). The atoms are labelled by the integer 
$n$ with $n = 0, 1, \cdots, N-1$. Hence, the cyclic character of 
the ring shape molecule makes it possible to identify the atom with 
the number $n$ to the one with the number $n+kN$ where $k \in \mathbb{Z}$ 
(the location of an atom is defined modulo $N$). Let 
$| \varphi_n \rangle$ be the atomic state vector, or atomic orbital 
in quantum chemistry parlance, describing a $\pi$--electron located 
in the neighboring of site $n$. From symmetry considerations, the 
molecular state vector, or molecular orbital, for the molecule 
reads$^{21}$
	\begin{eqnarray}
| \kappa_s \rangle = \frac{1}{\sqrt{N}} \sum_{n = 0}^{N-1} 
 {e} ^{{i} 2 \pi n s / N } | \varphi_n \rangle, 
	\label{molecular state vector}
	\end{eqnarray}
with $s = 0, 1, \cdots, N-1$. As a result, the molecular orbital  
$| \kappa_s \rangle$ assumes the same form, up to a global phase factor, 
as the state $| a \alpha \rangle$ given by 
Eq.~(\ref{a alpha in terms of k}) with $a=0$ and $\alpha = s$.

A similar result can be obtained for a one-dimensional chain of $N$ 
$1/2$--spins (numbered with $n=0, 1, \cdots, N-1$) used as a 
modeling tool of a ferromagnetic substance. Here again, we have a cyclical 
symmetry since the spins numbered $n=N$ and $n=0$ are considered to be 
identical. The spin waves can then be described by state 
vectors$^{21}$ very similar to the ones given by 
Eq.~(\ref{a alpha in terms of k}) with again $a=0$.  

We close this work with two remarks of a group-theoretical nature, one concerning 
a continuous group, the other a finite group, connected with the operators $u_{ab}$.

First, as mentioned in Section 4, the set 
$\{ u_{ab} : a, b = 0, 1, \cdots, d-1 \} \setminus \{ u_{00} \}$  constitutes a basis for the Lie algebra 
$SU(d)$. Such a basis differs from the well-known Cartan basis or from the Gel'fand-Tsetlin basis. In the 
special case $d=p$, with $p$ prime integer, the basis 
$\{ u_{ab} : a, b = 0, 1, \cdots, d-1 \} \setminus \{ u_{00} \}$ can be partioned into $p+1$ disjoint subsets,  
each subset containing $p-1$ commuting operators$^{18, 22}$. In other words, it is possible to decompose the 
Lie algebra of $SU(p)$ into $p+1$ Cartan subalgebras of dimension $p-1$. It can be proved that each subalgebra 
is associated with a basis of ${\cal E}(p)$ and that the set of the $p+1$ corresponding bases is a complete set 
of mutually unbiased bases. A similar decomposition holds of $SU(d)$ in the case where $d = p^e$, with $p$ prime 
integer and $e$ positive integer$^{22}$. However, in this case we need to replae ${\cal E}(d)$ by 
${\cal E}(p)^{\otimes e}$. 

A second group-theoretical remark concern a finite group known as the Pauli group or the finite 
Heisenberg-Weyl group$^{17, 18, 22, 23, 24}$. The set 
$\{ u_{ab} : a, b = 0, 1, \cdots, d-1 \}$ is not closed under multiplication.    
However, it is possible to extend the latter set in order to have a group. For 
this purpose, let us define the operators $w_{abc}$ via$^{22}$
	\begin{eqnarray}
w_{abc} = q^a u_{bc}, \quad a, b, c = 0, 1, \cdots, d-1. 
	\label{definition of wabc}
	\end{eqnarray}
Then, the set $\{ w_{abc} : a, b = 0, 1, \cdots, d-1 \}$, endowed with the multiplication of operators, is 
a group of order $d^3$. This group (the Pauli group) is of paramount importance in quantum information and 
quantum computation$^{24, 25}$. 

\newpage

\baselineskip = 0.60 true cm


\begin{thebibliography}{99}
\itemsep=-3pt



\bibitem{1Melvin} 
Melvin M.A.: {\it Rev. Mod. Phys.} {\bf 1956}, 28, 18.



\bibitem{2Racah} 
Racah G.: {\it Phys. Rev.} {\bf 1949}, 76, 1352. 



\bibitem{3Paldus} 
Paldus J., \v{C}\'{\i}\v{z}ek J., I. Shavitt: {\it Phys. Rev. A} {\bf 1972}, 5, 50; 

Paldus J.: {\it J. Chem. Phys.} {\bf 1974}, 57, 638;

Paldus J.: {\it J. Chem. Phys.} {\bf 1974}, 61, 5321;

Paldus J.: {\it Int. J. Quantum Chem. S} {\bf 1975}, 9, 165;

Paldus J., \v{C}\'{\i}\v{z}ek J.: {\it Adv. Quantum Chem.} {\bf 1975}, 9, 105;

Paldus J.: {\it Phys. Rev. A} {\bf 1976}, 14, 1620;

Paldus J.: {\it J. Chem. Phys.} {\bf 1977}, 67, 303;

Paldus J., Adams B. G., \v{C}\'{\i}\v{z}ek J.: {\it Int. J. Quantum Chem.} {\bf 1977}, 11, 813; 

Adams B. G., Paldus J., \v{C}\'{\i}\v{z}ek J.: {\it Int. J. Quantum Chem.} {\bf 1977}, 11, 849; 

Paldus J., \v{C}\'{\i}\v{z}ek J., Saute M., Laforgue A.: {\it Phys. Rev. A} {\bf 1978}, 17, 805;

Wormer P. E. S., Paldus J.: {\it Int. J. Quantum Chem.} {\bf 1979}, 16, 1307;

Paldus J., Wormer P. E. S.: {\it Int. J. Quantum Chem.} {\bf 1979}, 16, 1321;

Wormer P. E. S., Paldus J.: {\it Int. J. Quantum Chem.} {\bf 1980}, 18, 841;

Paldus J., Boyle M. J.: {\it Int. J. Quantum Chem.} {\bf 1982}, 22, 1281;

Paldus J., Takahashi M., Cho R. W. H.: {\it Phys. Rev. B} {\bf 1984}, 30, 4267;

Paldus J., Piecuch P.: {\it Int. J. Quantum Chem.} {\bf 1992}, 42, 135;

Li X., Paldus J.: {\it J. Chem. Phys.} {\bf 2003}, 119, 5334; 

Stuber J. L., Paldus J. in: {\it Fundamental World of Quantum Chemistry} 
(E.J. Br\"andas and E.S. Kryachko, Eds), Vol.~I. Kluwer, Dordrecht 2003. 



\bibitem{4Kibler} 
Grenet G., Kibler M.: {\it Phys. Lett. A} {\bf 1978}, 68, 147; 

Grenet G., Kibler M.: {\it Phys. Lett. A} {\bf 1979}, 71, 323; 

Kibler M., Grenet G.: {\it J. Math. Phys.} {\bf 1980}, 21, 422.



\bibitem{5Altmann} 
Altmann S. L., Cracknell A. P.: {\it Rev. Mod. Phys.} {\bf 1965}, 37, 19; 

Altmann S. L., Bradley C. J.: {\it Rev. Mod. Phys.} {\bf 1965}, 37, 33.



\bibitem{6Kibler} 
Kibler M.: {\it C.~R. Acad. Sci. (Paris) B} {\bf 1969}, 268, 1221; 

Kibler M. R.: {\it J. Math. Phys.} {\bf 1976}, 17,  855; 

Kibler M. R.: {\it J. Mol. Spectrosc.} {\bf 1976}, 62,  247; 

Kibler M. R.: {\it J. Phys. A: Math. Gen.} {\bf 1977}, 10, 2041; 

Kibler M. R., Guichon P. A. M.: {\it Int. J. Quantum Chem.} {\bf 1976}, 10, 87; 

Kibler M. R., Grenet G.: {\it Int. J. Quantum Chem.} {\bf 1977}, 11, 359; 

Kibler M. R.: {\it Int. J. Quantum Chem.} {\bf 1983}, 23, 115. 



\bibitem{7Moret}
Moret-Bailly J.: {\it J. Mol. Spectrosc.} {\bf 1965}, 15, 344; 

Michelot F., Moret-Bailly J.: {\it J. Phys. (Paris)} {\bf 1975}, 36, 451; 

Champion J. P., Pierre G., Michelot F., Moret-Bailly J.: {\it Can. J. Phys.} {\bf 1977}, 55, 512; 

Champion J. P.: {\it Can. J. Phys.} {\bf 1977}, 55, 1802. 


\bibitem{8PatWin}
Patera J., Winternitz P.: {\it J. Math. Phys.} {\bf 1973}, 14, 1130; 

Patera J., Winternitz P.: {\it J. Chem. Phys.} {\bf 1976}, 65, 2725. 



\bibitem{9Michel}
Michel L. in: {\it Group Theoretical Methods in Physics} 
(R. T. Sharp and B. Kolman, Eds). Academic Press, New York 1977. 



\bibitem{10Kibler}
Kibler M.: {\it J. Mol. Spectrosc.} {\bf 1968}, 26, 111; 

Kibler M.: {\it Int. J. Quantum Chem.} {\bf 1969}, 3, 795; 

Kibler M., Grenet G.: {\it Int. J. Quantum Chem.} {\bf 1985}, 28, 213; 

Kibler M., Grenet G.: {\it Int. J. Quantum Chem.} {\bf 1986}, 29, 11; 

Kibler M., Grenet G.: {\it Int. J. Quantum Chem.} {\bf 1986}, 29, 485; 

Kibler M., G\^acon J. C.: {\it Croat. Chem. Acta} {\bf 1989}, 62, 783.



\bibitem{11Edmonds} 
Edmonds A. R.: {\it Angular Momentum in Quantum Mechanics}, 
Princeton University Press, Princeton 1960.



\bibitem{12Kibler}
Kibler M. R.: {\it Collect. Czech. Chem. Commun.} {\bf 2005}, 70, 771;

Kibler M. R.: {\it Internat. J. Modern Phys. B} {\bf 2006}, 20, 1792.



\bibitem{13KibPlaAlbKibKib}
Kibler M. R., Planat M.: {\it Int. J. Mod. Phys. B} {\bf 2006}, 20, 1802;

Albouy O., Kibler M. R.: {\it SIGMA} {\bf 2007} 3, article 076;

Kibler M. R.: {\it SIGMA} {\bf 2007}, 3, article 092. 



\bibitem{14mubs}
Ivanovi\'c I. D.: {\it J. Phys. A: Math. Gen.} {\bf 1981}, 14, 3241;

Wootters W. K., Fields B. D.: {\it Ann. Phys. (N Y)} {\bf 1989}, 191, 363;

Calderbank A. R., Cameron P. J., Kantor W. M., Seidel J. J.: {\it Proc. London Math. Soc.} {\bf 1997}, 75, 436;

Bandyopadhyay S., Boykin P. O., Roychowdhury V., Vatan F.: {\it Algorithmica} {\bf 2002}, 34, 512;

Lawrence J., Brukner \v{C}., Zeilinger A.: {\it Phys. Rev. A} {\bf 2002}, 65, 032320;

Lawrence J.: {\it Phys. Rev. A} {\bf 2004}, 70, 012302;

Klappenecker A., R\"otteler M.: {\it Lect. Notes Comput. Sci.} {\bf 2004}, 2948, 137;

Gibbons K. S., Hoffman M. J., Wootters W. K.: {\it Phys. Rev. A} {\bf 2004}, 70, 062101; 

Pittenger A. O., Rubin M. H.: {\it Linear Algebr. Appl.} {\bf 2004}, 390, 255;

Pittenger A. O., Rubin M. H.: {\it J. Phys. A: Math. Gen.} {\bf 2005}, 38, 6005. 



\bibitem{15Weyl}
Weyl H.: {\it The Theory of Groups and Quantum Mechanics}, Dover Publications, New York 1931.



\bibitem{16McIntosh}
McIntosh H. V.: {\it J. Molec. Spectrosc.} {\bf 1962}, 8, 169.



\bibitem{17Balian}
Balian R., Itzykson C.: {\it C. R. Acad. Sci. (Paris)} {\bf 1986}, 303, 773. 



\bibitem{18PateraZassenhaus}
Patera J., Zassenhaus H.: {\it J. Math. Phys.} {\bf 1988}, 29, 665.



\bibitem{19autres}
Galetti D., De Toledo Piza A. F. R.: {\it Physica A} {\bf 1988}, 149, 267;

Knill E.: {\it arXiv} {\bf 1996}, quant-ph/9608048.

Gottesman D.: {\it Chaos, Solitons and Fractals} {\bf 1999}, 10, 1749;

Pittenger A. O., Rubin M. H.: {\it Phys. Rev. A} {\bf 2000}, 62, 032313; 

Gottesman D., Kitaev A., Preskill J.: {\it Phys. Rev. A} {\bf 2001}, 64, 012310;

Bartlett S. D., de Guise H., Sanders B. C. : {\it Phys. Rev. A} {\bf 2002}, 65, 052316;

Klimov A. B., S\'anchez-Soto L. L., de Guise H.: {\it J. Phys. A: Math. Gen.} {\bf 2005}, 38, 2747.



\bibitem{20Gell-MannNe'eman}
Gell-Mann M.: {\it Phys. Rev.} {\bf 1962}, 125, 1067; 

Ne'eman Y.: {\it Nucl. Phys.} {\bf 1961}, 26, 222. 



\bibitem{21LeBellac}
Le Bellac M.: {\it Physique quantique}, EDP Sciences/CNRS \'Editions, Paris 2003.



\bibitem{22progress}
Kibler M. R.: work in progress. 



\bibitem{23Wolf}
Wolf K. B., Garc\'{\i}a A.: {\it Rev. Mex. F\'\i sica} {\bf 1972}, 21, 191;

Wolf K. B. in: {\it Group theory and its applications} (E. M. Loebl, Ed), Vol.~III. Academic Press, New York 1975.



\bibitem{24reste}
Grassl M.: {\it Elec. Notes Discrete Math.} {\bf 2005}, 20, 151;

Durt T.: {\it J. Phys. A: Math. Gen.} {\bf 2005}, 38, 5267;

Appleby D. M.: {\it J. Math. Phys.} {\bf 2005}, 46, 052107;

Flammia S. T.: {\it J. Phys. A: Math. Gen.} {\bf 2006}, 39, 13483;

Cormick C., Galv\~ao E. F., Gottesman D., Paz J. P., Pittenger A. O.: 
{\it Phys. Rev. A} {\bf 2006}, 73, 012301; 

Vourdas A.: {\it J. Phys. A: Math. Theor.} {\bf 2007}, 40, R285.



\bibitem{25econet} 
Planat M., Saniga M., Kibler M. R.: {\it SIGMA} {\bf 2006}, 2, paper 066;

Havlicek H., Saniga M.: {\it J. Phys. A: Math. Theor.} {\bf 2007}, 40, F943;

Planat M., Baboin A.-C.: {\it J. Phys. A: Math. Theor.} {\bf 2007}, 40, F1005; 

Planat M., Saniga M.: {\it Quantum Inf. Comput.} {\bf 2008}, 8, 0127;

Planat M., Baboin A.-C., Saniga M.: {\it Int. J. Theor. Phys.} {\bf 2008}, 47, 1127;

Havlicek H., Saniga M.: {\it J. Phys. A: Math. Theor.} {\bf 2008}, 41, 015302; 

Planat M., Jorrand P.: {\it J. Phys. A: Math. Theor.} {\bf 2008} 41, 182001. 

\end{thebibliography}
\end{document}